\documentclass[12pt,epsf,amstex]{article}
\usepackage [dvips]{graphicx}
\usepackage{amsmath}
\usepackage{amssymb}
\usepackage{epsfig}

\addtocounter{secnumdepth}{1}
\begin{document}

\newcommand{\bP}{\bar{P}}
\newcommand{\br}{\bar{r}}
\newcommand{\bv}{\bar{v}}
\newcommand{\bz}{\bar{z}}
\newcommand{\ac}{a_{\rm c}}
\newcommand{\bc}{b_{\rm c}}
\newcommand{\Bc}{B_{\rm c}}
\newcommand{\Kc}{K_{\rm c}}
\newcommand{\Tc}{T_{\rm c}}
\newcommand{\apm}{a_{\pm}}
\newcommand{\bpm}{b_{\pm}}

\newcommand{\ms}{m_{\rm sp}}
\newcommand{\bbeta}{\bar{\beta}}
\newcommand{\bgamma}{\bar{\gamma}}
\newcommand{\btau}{\bar{\tau}}
\newcommand{\bomega}{\bar{\omega}}
\newcommand{\bOmega}{\bar{\Omega}}

\newcommand{\tbeta}{\tilde{\beta}}
\newcommand{\tgamma}{\tilde{\gamma}}
\newcommand{\trho}{\tilde{\rho}}
\newcommand{\ttau}{\tilde{\tau}}
\newcommand{\tLambda}{\tilde{\Lambda}}
\newcommand{\tOmega}{\tilde{\Omega}}
\newcommand{\td}{\tilde{d}}
\newcommand{\tN}{\tilde{N}}
\newcommand{\tT}{\tilde{T}}
\newcommand{\tm}{\tilde{m}}
\newcommand{\tchi}{\tilde{\chi}}

\newcommand{\hrho}{\hat{\rho}}
\newcommand{\htau}{\hat{\tau}}

\newcommand{\bE}{{\bf{E}}}
\newcommand{\bO}{{\bf{O}}}
\newcommand{\bR}{{\bf{R}}}
\newcommand{\bS}{{\bf{S}}}
\newcommand{\bT}{\mbox{\bf T}}
\newcommand{\bt}{\mbox{\bf t}}
\newcommand{\half}{\frac{1}{2}}
\newcommand{\thalf}{\tfrac{1}{2}}
\newcommand{\bsA}{\mathbf{A}}
\newcommand{\bsV}{\mathbf{V}}
\newcommand{\bsE}{\mathbf{E}}
\newcommand{\bsT}{\mathbf{T}}
\newcommand{\bsZ}{\hat{\mathbf{Z}}}
\newcommand{\bse}{\mbox{\bf{1}}}
\newcommand{\bspsi}{\hat{\boldsymbol{\psi}}}
\newcommand{\cdottt}{\!\cdot\!}
\newcommand{\deltaR}{\delta\mspace{-1.5mu}R}
\newcommand{\invup}{\rule{0ex}{2ex}}

\newcommand{\bGamma}{\boldmath$\Gamma$\unboldmath}
\newcommand{\dd}{\mbox{d}}
\newcommand{\ee}{\mbox{e}}
\newcommand{\p}{\partial}

\newcommand{\rmax}{r_{\rm max}}

\newcommand{\artanh}{\mbox{artanh}}

\newcommand{\wrj}{w^{r}_j}
\newcommand{\wrzerj}{w^{r}_{0,j}}
\newcommand{\wronej}{w^{r}_{1,j}}
\newcommand{\wrtwoj}{w^{r}_{2,j}}
\newcommand{\wsj}{w^{s}_j}
\newcommand{\Wr}{W^{\rm r}}
\newcommand{\Ws}{W^{\rm s}}
\newcommand{\Wrj}{W^{r}_j}
\newcommand{\Wsj}{W^{s}_j}
\newcommand{\Wsi}{W^{s}_i}
\newcommand{\wstarrj}{w^r_{*,j}}
\newcommand{\wstarsj}{w^s_{*,j}}

\newcommand{\wGj}{w^{\rm G}_j}
\newcommand{\Pst}{P_{\rm st}}
\newcommand{\Pstzero}{P_{{\rm st},0}}
\newcommand{\Pstone}{P_{{\rm st},1}}
\newcommand{\Pstupone}{P_{\rm st}^{(1)}}
\newcommand{\Zstupone}{Z^{(1)}}
\newcommand{\Zstell}{Z_{\ell}^{(1)}}
\newcommand{\calWupone}{{\cal W}^{(1)}}
\newcommand{\calHupone}{ {\cal H}^{(1)} }
\newcommand{\raupone}{ \rangle^{(1)} }
\newcommand{\gupone}{g^{(1)}}
\newcommand{\tgupone}{\tilde{g}^{(1)}}
\newcommand{\wupone}{w^{(1)}}
\newcommand{\wjtwo}{w_{j,2}}
\newcommand{\wjzero}{w_{j,0}}
\newcommand{\raG}{\rangle^{(1)}_{\rm G}}

\newcommand{\la}{\langle}
\newcommand{\ra}{\rangle}
\newcommand{\rao}{\rangle\raisebox{-.5ex}{$\!{}_0$}}  
\newcommand{\rae}{\rangle\raisebox{-.5ex}{$\!{}_1$}}
\newcommand{\beq}{\begin{equation}}
\newcommand{\eeq}{\end{equation}}
\newcommand{\bea}{\begin{eqnarray}}
\newcommand{\eea}{\end{eqnarray}}
\def\lsim{\:\raisebox{-0.5ex}{$\stackrel{\textstyle<}{\sim}$}\:}
\def\gsim{\:\raisebox{-0.5ex}{$\stackrel{\textstyle>}{\sim}$}\:}

\numberwithin{equation}{section}

\thispagestyle{empty}
\title{\Large {\bf 
Two interacting Ising chains\\[2mm]
in relative motion
}} 
 
\author{{H.J.~Hilhorst}\\[5mm]
{\small Laboratoire de Physique Th\'eorique, B\^atiment 210}\\[-1mm] 
{\small Universit\'e Paris-Sud and CNRS,
91405 Orsay Cedex, France}\\}

\maketitle
\begin{small}
\begin{abstract}
\noindent 
We consider
two parallel cyclic Ising chains counter-rotating
at a relative velocity $v$,
the motion actually being a succession of discrete steps.
There is an in-chain interaction 
between nearest-neighbor spins and a  
cross-chain interaction
between instantaneously opposite spins.
For velocities $v > 0$ the system, subject to
a suitable markovian dynamics at a temperature $T$,
can reach only a nonequilibrium steady state (NESS). 
This system was introduced
by Hucht et al., who showed that for $v=\infty$ 
it undergoes a para- to ferromagnetic
transition, essentially due to the fact that each chain exerts an
effective field on the other one.
The present study of the $v=\infty$ case
determines the consequences of the fluctuations of this effective field
when the system size N is finite.
We show that whereas to leading order the system obeys
detailed balancing with respect to an
effective time-independent Hamiltonian, the higher order
finite-size corrections violate detailed balancing.
Expressions are given to various orders in 
$N^{-1}$ for the interaction free energy between the chains, 
the spontaneous magnetization, the in-chain and cross-chain
spin-spin correlations, and the spontaneous magnetization.
It is shown how finite-size
scaling functions may be derived explicitly.
This study was motivated by 
recent work on a two-lane traffic problem
in which a similar phase transition was found.\\

\end{abstract}
\end{small}
\vspace{12mm}

\noindent LPT Orsay 11/03


\section{Introduction} 
\label{secintroduction}

Recently Hucht \cite{Hucht09} (see also \cite{Kadauetal08}),
motivated by the phenomenon of magnetic friction,
formulated a nonequilibrium steady state (NESS) Ising model of a new type.
It consists of two parallel linear Ising chains having a relative
velocity $v$.
In addition to a nearest-neighbor interaction in each chain, 
any pair of spins facing each other 
on the two chains has an instantaneous interaction.
In the version of the model easiest to study, each chain is finite and
periodic; we will therefore speak of
cyclic counter-rotating Ising chains (CRIC).
The model, subject to suitable temperature dependent
Markovian dynamics,  
was shown \cite{Hucht09} at velocity $v=\infty$
to have a para- to ferromagnetic phase transition 
which in the limit of infinitely long chains may be
understood in terms of an equivalent equilibrium model.

The CRIC seems to us to be of the same fundamental importance 
as Glauber's
\cite{Glauber63} original kinetic Ising model.
First, it is of interest in its own right as a new member
of the class of NESS.
Second, its interest is enhanced in the wider context of recent work on Ising
models that in one way or another are driven,
dissipate energy, or have some novel type of coupling;
such work has appeared
in a variety of contexts \cite{DemeryDean10,Pleimlingetal10,Pradosetal10}.
In particular, the present CRIC was extended to a Potts version by
Igl\'oi {\it et al.} \cite{Igloietal11}, who find remarkable
nonequilibrium phase transitions. 
In this paper we contribute further to the study of the CRIC.
We focus on finite chains and on how to derive
known and new properties from the master equation that defines the model. 
\vspace{3mm}

Hucht's solution \cite{Hucht09} is based on showing that
at $v=\infty$ the stationary state dynamics of
the CRIC is actually that of an equilibrium Ising chain 
in an effective magnetic field $H_0$, 
this field being zero above the transition
temperature and nonzero below. This equivalence 
is valid in the limit where the chain length $N$ tends to infinity.
In this work we show 
that it is possible to formulate this problem as an expansion
in powers of $N^{-1/2}$. To lowest order we recover the equivalent
equilibrium system found in reference \cite{Hucht09}.
To higher orders fluctuations of the field $H_0$ come into play and 
appear as finite-size effects.

The finite $N$ case is of interest, first of all, on the level of principles,
and secondly, for the analysis of finite size effects in simulations
as were carried out in \cite{Hucht09} and by ourselves.
We expect, furthermore, that our approach will help prepare the way
for future work on the $v<\infty$ case, which is considerably harder.

The effective transition rates satisfy detailed balancing
to leading order in the large-$N$ expansion \cite{Hucht09};
our analysis reveals, however, that
to higher orders in $N^{-1/2}$ the
detailed balancing (DB) symmetry of the effective rates is broken. 
The stationary state distribution may be found explicitly,
at least to the lowest DB-violating order. 
Knowing this state one can calculate all desired NESS properties.
\vspace{5mm}


In section \ref{secdefmodel} of this paper
we define the rules of the markovian dynamics for general relative 
velocity $v$ and then specialize to $v=\infty$. These dynamical
equations are the starting point for all that follows.
In section \ref{secdetbal} we discuss the DB violation 
that occurs in higher orders of $N^{-1}$.
In section \ref{seczero} we consider the stationary state to zeroth
order, as was already done by Hucht \cite{Hucht09}. 
In sections \ref{secexpansion} and \ref{secfirst}
we show how $N^{-1}$ can be introduced
as an expansion parameter and we define a `leading order', composed of
the zeroth order and a first-order correction.
In section \ref{secsecond} we show how for the stationary state
distribution an expansion may be found in powers of $N^{-1}$. We
present the explicit result to next-to-leading order.
In section \ref{secaverages} we calculate for various quantities of
physical interest their stationary state
averages to successive orders in the expansion. 
In section \ref{sectraffic} we briefly discuss the relation of the
present model to a two-lane road traffic model studied earlier. 
In section \ref{secconclusion} we conclude.


\section{Counter-rotating Ising chains}
\label{secdefmodel}


\subsection{A stochastic dynamical system}
\label{secmodel}

\begin{figure}
\begin{center}
\scalebox{.55}
{\includegraphics{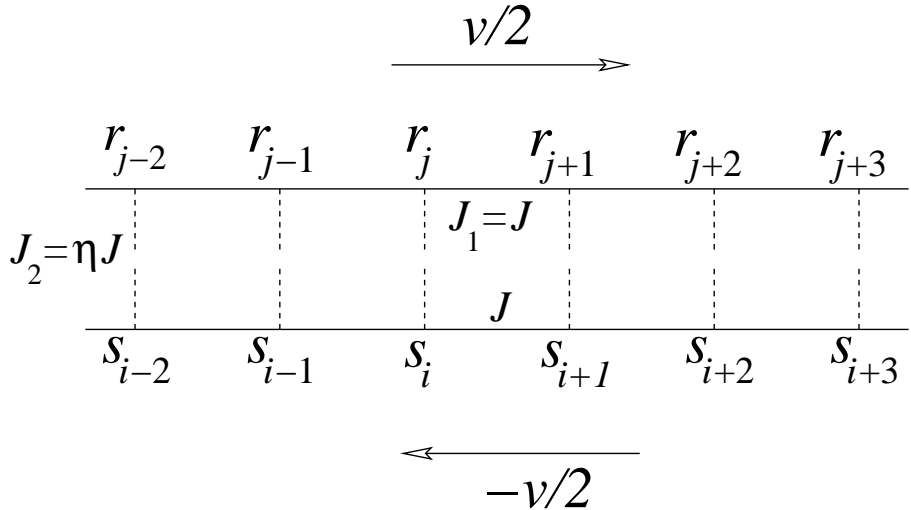}}
\end{center}
\caption{\small Ladder of spins with an intrachain nearest-neighbor
interaction $J_1=J$.
The two chains constituting the ladder
have a relative velocity $v$, the motion taking place in
discrete steps of one lattice unit. 
There is an interchain nearest-neighbor
interaction $J_2=\eta J$ between each pair of spins facing each other
at any instant in opposite chains.} 
\label{figladder}
\end{figure}

We consider Ising spins on the ladder lattice shown in figure \ref{figladder}.
The spins in the upper chain are denoted by $r_j$, those in the lower chain by
$s_i$, where the integers $j$ and $i$ are site indices.
There is a nearest-neighbor interaction
$J_1=J$ inside each chain 
and an interaction $J_2=\eta J$ between each pair of spins facing each other
in opposite chains. We take $J>0$ and $\eta$ of arbitrary sign.
The feature \cite{Kadauetal08} and \cite{Hucht09} 
that distinguishes this model from the standard Ising model on a
ladder lattice,
is that the two chains move with respect to one another at a speed $v>0$. 
This will mean the following:
the time axis is discretized in intervals
of duration $\tau=a_0/v$ (where $a_0$ is the lattice spacing) 
and at the end of each interval the upper 
chain is shifted one lattice spacing $a_0$
to the right with respect to the lower one. 
The Hamiltonian ${\cal H}(t)$ of this system
is therefore time-dependent and given by%
\beq
{\cal H}(t) = -J\sum_j \left[ r_jr_{j+1} + s_js_{j+1} \right]  
-\eta J \sum_j r_{j}s_{\lfloor j+vt/a_0 \rfloor}\,,
\label{Hamiltonian}
\eeq
where $\lfloor x \rfloor$ is the largest integer less than or
equal to $x$.

We will consider cyclic boundary conditions%
\footnote{In connection with the traffic problem
open boundary conditions
are certainly also worthy of consideration. These have however
the inconvenience of breaking the translational symmetry.}. 
In this case the chains become counter-rotating
loops of length say $N$;
the site indices $i$, $j$, and
$\lfloor j+vt/a_0 \rfloor$ must then be interpreted modulo $N$.
Employing
the shorthand notation $r=\{r_j|j=1,2,\ldots,N\}$
and $s=\{s_j|j=1,2,\ldots,N\}$, we may indicate
a spin configuration of the system by $(r,s)$.

We associate with ${\cal H}(t)$ a stochastic
time evolution of $(r,s)$. Its precise definition requires that
we exercise some caution.
We will first define it as a Monte Carlo procedure and then write down
the master equation and pass to analytic considerations.
Single-spin reversals are attempted
at uniformly distributed random instants of time
at a rate of $1/\tau_0$ per site%
\footnote{We may scale time such that $\tau_0=1$.}.
Each attempt 
is governed by transition probabilities.
Since there are $2N$ sites, there are $2N$ different single-spin flips
by which a state $(r,s)$ may be entered or exited. 
Given that a reversal attempt takes place, let 
$(2N)^{-1}\Wr_j(r;s;t)$ and $(2N)^{-1}\Ws_j(s;r;t)$ be the probabilities that
$r_j$ and $s_j$ are flipped, respectively.
The reversal attempt will remain unsuccessful with the
complementary probability
\beq
1-A_{\rm acc} = 1-(2N)^{-1}\sum_{j=1}^N 
\left[ \Wr_j(r;s;t) + \Ws_j(s;r;t) \right],
\label{defAacc}
\eeq
where $A_{\rm acc}$ is what is usually called the `acceptance
probability'. 

We now specify the $\Wr_j$ and $\Wr_j$ in such a way that
at any time $t$ the system strives to attain the canonical equilibrium
at a given temperature $T$ with respect to the
instantaneous Hamiltonian ${\cal H}(t)$.
The choice is not unique.
We choose
\bea
\Wr_j(r;s;t)&=&\tfrac{1}{4}
\big[ 1-\tfrac{1}{2}r_j(r_{j-1}+r_{j+1})\tanh 2K \big]
\big[ 1-r_js_i\tanh\eta K \big], \nonumber\\[2mm]
\Ws_i(s;r;t)&=&\tfrac{1}{4}
\big[ 1-\tfrac{1}{2}s_i(s_{i-1}+s_{i+1})\tanh 2K \big]
\big[ 1-s_ir_j\tanh\eta K \big],
\label{defw}
\eea
where we have set $K=J/T$
(with $T$ measured in units of Boltzmann's constant)
and where in {\it both\,} equations $i$ and $j$ are related by
\beq
i = \lfloor j+vt/a_0 \rfloor\!\!\mod N.
\label{relijt}
\eeq
Equation (\ref{defw}) is different both from the heat bath (or: Glauber)
and from the Metropolis transition probabilities.
We will refer to it as the ``factorizing rate''.
The factor
\beq
\wGj(r)=\tfrac{1}{2}\big[ 1-\tfrac{1}{2}r_j(r_{j-1}+r_{j+1})\tanh 2K \big]
\label{defwG}
\eeq
represents the Glauber transition probability.
The $\Wr_j$ and $\Ws_j$ define an easy-to-simulate Markov chain%
\footnote{No confusion should arise with the two 
  legs of the ladder lattice, to which we refer also as `chains'.}
with time-dependent transition probabilities.%
\footnote{The reversal attempts, that is, the steps of the Markov
  chain, are Poisson distributed on the time axis.
This makes it possible at any time to probabilistically
connect the elapsed time $t$ to the
number of spin reversal attempts $n$. In the large $t$ limit
of course $n\simeq t/\tau_0$.}
 
In the special case
$v=0$ the Hamiltonian ${\cal H}(t)$ reduces to the equilibrium 
Hamiltonian of the ladder lattice. For $v$ arbitrary but $\eta=0$
it reduces to the equilibrium Hamiltonian of two decoupled chains. In
both special cases the dynamics is standard and obeys detailed balancing.

In the general case, since
the Hamiltonian is time-dependent, the system will not
reach equilibrium but instead enter a NESS. 
Actually, for generic $v$,
because of the periodic discrete shifts, the NESS is a
$\tau$-periodic function of time;
NESS averages are naturally defined to include an average over this period.
In the limiting case
$v=\infty$ we have $\tau=0$ and this complication disappears.
The infinite velocity
NESS is the subject of our interest in the remaining sections.
It is a problem that depends only on the two parameters $K$ and $\eta$.

We note finally that as compared to ours, there is an extra prefactor
\beq
\frac{2+(1-r_{j-1}r_{j+1})\tanh 2K}{1+\tanh 2K}(1+\ee^{-2\eta K})
\label{prefactor}
\eeq
in Hucht's expression for the transition probability $\Wr_j(r;s;t)$,
and an analogous prefactor for $\Ws_i(s;r;t)$.
These factors may easily be carried along in the calculation.


\subsection{The limit $v\to\infty$}
\label{secvinfty}

Let $P(r,s;n)$ be the probability distribution on the configurations
$(r,s)$ after $n$ spin reversal attempts. 
We will now write down the formal evolution equation for
$P(r,s;n)$ for the case of $v=\infty$, 
where important simplifications occur.
When $v=\infty$ there is no relation between the indices $i$ and $j$
and hence the chain has transition probabilities
$w_j(r;s)$ given by the average of (\ref{defw})
on all $i$, which is now considered as an independent variable.  
We denote this average by $w_j(r,s)$ and thus have 
\begin{subequations}\label{defwrsj}
\bea
w_j(r;s) &=& \frac{1}{N}\sum_{i=1}^N \Wr_j(r;s;t) \nonumber\\[2mm]
&=& \tfrac{1}{4}\big[ 1-\tfrac{1}{2}r_j(r_{j-1}+r_{j+1})\tanh 2K \big]
\big[ 1-r_j\mu(s)\tanh\eta K\big] \nonumber\\[2mm]
&=& \wGj(r)\times\tfrac{1}{2}\big[ 1-r_j\mu(s)\tanh\eta K\big], 
\label{defwrj}\\[2mm]
w_j(s;r) &=& \frac{1}{N}\sum_{i=1}^N \Ws_j(s;r;t) \nonumber\\[2mm]
&=&\wGj(s)\times\tfrac{1}{2}\big[ 1-s_j\mu(r)\tanh\eta K\big], 
\label{defwsj}
\eea
\end{subequations}
where 
\beq
\mu(s)=\frac{1}{N}\sum_{i=1}^Ns_i, \qquad \mu(r)=\frac{1}{N}\sum_{i=1}^Nr_i\,.
\label{defmus}
\eeq
We will write $r^j$ for the configuration
obtained from $r$ by reversing $r_j$
(that is, by carrying out the replacement $r_j \mapsto -r_j$), 
and similarly define $s^j$.
Summing on all $2N$ flips by which it is possible to enter or to exit
$(r,s)$ we find that the evolution of $P(r,s;n)$ is described by the
master equation
\bea
P(r,s;n+1) &=& \frac{1}{2N}\sum_{j=1}^N \left[\, w_j(r^j;s)P(r^j,s;n)
 + w_j(s^j;r)P(r,s^j;n) \right.\nonumber\\[2mm]
&&  + \left. \big( 1-w_j(r;s) \big)P(r,s;n) 
    + \big( 1-w_j(s;r) \big)P(r,s;n) \,\right],
\nonumber\\
&&
{}\label{defMchain}
\eea
where the second line corresponds to the probability of an unsuccessful spin
reversal attempt. In vector notation equation
(\ref{defMchain}) may be written
\beq
P(n+1) = ({\bf 1} + {\cal W}) P(n),
\label{defW}
\eeq
where $P(n)$ is the $2^{2N}$-dimensional vector of elements 
$P(r,s;n)$, the symbol ${\bf 1}$ denotes the unit matrix, and
${\cal W}$ is a matrix composed of entries $w_j$ for
which comparison of (\ref{defMchain}) 
and (\ref{defW}) yields
\bea
{\cal W}(r,s;r',s')&=&\delta_{r'r^j}\delta_{s's}w_j(r^j,s)
                     +\delta_{r'r}\delta_{s's^j}w_j(r,s^j) \nonumber\\[2mm]
&&-\,\delta_{r'r}\delta_{s's}\sum_{j=1}^N \left[ w_j(r;s)+w_j(s;r) \right].
\label{relwW}
\eea
The discrete-time master equation,
(\ref{defMchain}) together with the
Poisson statistics of the reversal attempts on the time axis, 
fully defines the CRIC for $v=\infty$.
This equation may be studied analytically, as is
the purpose of this work, or may be implemented in a Monte Carlo
simulation. 


\section{Detailed balancing and its violation}
\label{secdetbal}

Henceforth we consider the case $v=\infty$. 
Our purpose is now to find the 
stationary state distribution $P_{\rm st}(r,s)$ of the evolution
equation (\ref{defMchain}).
This distribution 
is the solution of $P(r,s;n)=P(r,s;n+1)=P_{\rm st}(r,s)$, which means
\beq
0={\cal W}P_{\rm st}\,.
\label{defstst}
\eeq
Combining equations (\ref{defstst}) and (\ref{defMchain})
yields the $v=\infty$ stationary state equation
\bea 
0 &=& \sum_{j=1}^N \left[w_j(r^j;s)\Pst(r^j,s) 
 + w_j(s^j;r)\Pst(r,s^j) \right. \nonumber\\
&&\phantom{XXXX}  
- \left. w_j(r;s)\Pst(r,s) - w_j(s;r)\Pst(r,s) 
\right]. 
\label{defPst}
\eea
If the transition probabilities
satisfy the condition
of detailed balancing, the solution of (\ref{defPst}) is easily constructed;
in case of the contrary, there are no general methods.
We examine therefore first the question of whether equation (\ref{defMchain})
satisfies detailed balancing.

A Markov chain satisfies detailed balancing (DB)
if and only if 
its transition probabilities are such that
any loop in configuration space is traversed with equal probability in 
either direction.
To show that the transition probabilities $w_j$ 
fail to obey DB
we consider an elementary
loop of four single-spin flips, 
\beq
(r,s) \mapsto (r^j,s) \mapsto (r^j,s^j) \mapsto (r,s^j) \mapsto (r,s).
\label{defloop}
\eeq
Given the system is in $(r,s)$, we denote by $p_+(\eta)$ and $p_-(\eta)$ 
the probability that in the next four attempts it goes
through this loop in forward and in backward direction, respectively.
That is,
\bea
p_+(\eta) &=& w_j(r;s)w_j(s;r^j)w_j(r^j;s^j)w_j(s^j;r), \nonumber\\[2mm]
p_-(\eta) &=& w_j(s;r)w_j(r;s^j)w_j(s^j;r^j)w_j(r^j;s).
\label{xppmeta}
\eea
For $\eta=0$ the two chains are decoupled, and as discussed below
equation (\ref{relijt}), each of them separately satisfies DB; 
it is easy indeed to verify explicitly that $p_+(0)=p_-(0)\equiv p(0)$. 
For general $\eta$ we may
work out the difference $p_+(\eta)-p_-(\eta)$ with the aid of (\ref{defwrj}),
(\ref{defmus}), and the relations
\beq
\mu(r^j) = \mu(r)-\frac{2r_j}{N}\,, \qquad \mu(s^j) = \mu(s)-\frac{2s_j}{N}\,, 
\label{relmurjmur}
\eeq
which yields
\bea
p_+(\eta)-p_-(\eta) &=& 4N^{-1}  p(0)\,\tanh^2\eta K\,
[r_j\mu(s)-s_j\mu(r)] \nonumber\\[2mm]
&& \times \big\{ [r_j\mu(r)-s_j\mu(s)]+2N^{-1}\tanh\eta K \big\}.
\label{xdiffppmeta}
\eea
This shows that DB is violated in the general case of
nonzero coupling ($\eta\neq 0$)
between the chains. It becomes valid again only asymptotically in
the limit $N\to\infty$.
We therefore cannot hope to rely on any general methods to construct
$P_{\rm st}(r,s)$ for finite $N$.
Indeed, writing out the stationary state equation
(\ref{defstst}) fully explicitly for $N=3,4$
(only $N=2$ is trivial) has confirmed the nontriviality of the
stationary state but has not provided us with any useful insight. 


\section{Stationary state $P_{\rm st}(r,s)$  to zeroth order}
\label{seczero}

The limit $N\to\infty$ was considered by Hucht \cite{Hucht09,Kadauetal08},
and we briefly recall the results. 
One may suppose that in this limit $\mu(r)$ and $\mu(s)$
have vanishing fluctuations around an
as yet unknown common average to be called $m_0(K,\eta)$. 
We will denote the $N\to\infty$ limit of $w_j$ by $w_{j,0}$\,.
It then follows from (\ref{defwrj}) that
\beq
w_{j,0}(r) = \wGj(r) \times \tfrac{1}{2}[1-r_jm_0\tanh\eta K].
\label{defwrj0}
\eeq
With the transition probabilities (\ref{defwrj0}) the $r$- and the
$s$-chain decouple. Moreover, the expression for these $w_{j,0}$
is such that the spin dynamics satisfies
DB with respect to the pair of uncoupled 
nearest-neighbor Ising Hamiltonians in a field,
\beq
{\cal H}_0(r,s)/{T} =  
- K\sum_{j=1}^{N} \big[ r_jr_{j+1} + s_js_{j+1} \big]
- H_0\sum_{j=1}^{N} \big[ r_j + s_j \big]\,.
\label{defH0}
\eeq
where $H_0$ is defined in terms of $m_0$ by
\beq
\tanh H_0 = m_0\tanh\eta K
\label{deffield}
\eeq
and where $K$ and $H_0$ both include a
factor $1/T$. 
The quantity ${\cal H}_0(r,s)$ is an effective time-independent Hamiltonian.
Let $m(K,z)$ denote the magnetization per spin of the one-dimensional (1D)
Ising chain with coupling $K$ in a field that we will for convenience
denote by $z$. This quantity is well-known and given by
\beq
m(K,z)=\frac{\sinh z}{\sqrt{ \sinh^2 z + {\rm e}^{-4K} }}\,.
\label{xmKz}
\eeq
Consistency requires that 
\beq
m_0=m(K,H_0).
\label{relconst}
\eeq 
Upon combining (\ref{deffield}) with (\ref{relconst})
one obtains an equation for $H_0$ [or equivalently $m_0$].
The solution $H_0$ is a function of the two system parameters $K$
and $\eta$ and given by
\beq
\tanh H_0(K,\eta)=\left\{
\begin{array}{ll}
\left( \dfrac{ \tanh^2\eta K-{\rm e}^{-4K} }
{ 1-{\rm e}^{-4K} } \right)^{\half}, \phantom{XX}& K>\Kc, \\[5mm]
0, & K\leq\Kc\,,
\end{array}
\right.
\label{solnH0}
\eeq
in which there appears a critical coupling  $\Kc = J/\Tc$ that is the
solution of%
\footnote{Equation (\ref{xKc}) may be rewritten as 
$\sinh(2J_1/\Tc)\sinh(2J_2/\Tc)=1$, which shows, as
  was also noticed in reference \cite{Hucht09}, that $\Tc$ is exactly 
  (but accidentally) equal to the critical
  temperature of Onsager's square Ising model with horizontal and
  vertical couplings $J_1$ and $J_2$.}
\beq
\tanh \eta \Kc = \ee^{-2\Kc}.
\label{xKc}
\eeq
The magnetization $m_0(K,\eta)$ follows directly from (\ref{deffield}) 
and (\ref{solnH0}). For $T\to\Tc^-$ it vanishes
as $m_0 \propto (\Tc-T)^{\beta}$ with a classical exponent $\beta=\frac{1}{2}$.
For later use it is worthwhile to notice that also 
$H_0(T)\propto (T-\Tc)^{1/2}$ when $T\leq\Tc$\,.

The DB property found below equation (\ref{defwrj0})
now allows us to conclude that for $N\to\infty$
the stationary state distribution $P_{{\rm st},0}(r,s)$
is the Boltzmann distribution corresponding to (\ref{defH0}), that is,
\beq
P_{{\rm st},0}(r,s) = {\cal N}_0\, \ee^{-{\cal H}_0(r,s)/T}
\label{xstst0}
\eeq
where ${\cal N}_0$ is the normalization.
In reference \cite{Hucht09} several
system properties were calculated in this $N\to\infty$
limit by averaging with respect to $P_{{\rm st},0}(r,s)$.


\section{Expansion procedure for $P_{\rm st}(r,s)$}
\label{secexpansion}

As has become clear in section \ref{secdetbal},
the inverse system size $1/N$ is a measure of the degree of DB
violation. 
In the present case this will lead us to attempt
to find the finite $N$ stationary state by expanding around
the known $N=\infty$ solution (\ref{xstst0}), which will
play the role of the zeroth order result.
At the basis of the expansion is the hypothesis, to be confirmed
self-con\-sistent\-ly, that the fluctuations 
$\delta\mu$ of the chain magnetizations, defined by
\beq
\delta\mu(r)=\mu(r)-m_0\,, \qquad \delta\mu(s)=\mu(s)-m_0\,.
\label{defmurs}
\eeq
are of order $N^{-1/2}$.
\vspace{3mm}

A naive attempt to set up the expansion
would be to notice that
the transition probability (\ref{defwrj})
can be written as a sum of its average and a correction, 
$w_j(r;s) = w_{j,0}(r) + \bar{w}_{j}(r;s)$,
where $w_{j,0}(r)$ is given by (\ref{defwrj0}) and
$\bar{w}_{j}(r;s) = \wGj(r) \times(-\tfrac{1}{2}r_j) \delta\mu(s) \tanh\eta K$
is of order $N^{-1/2}$. One might then
think that there exists a corresponding expansion
$P_{\rm st}(r,s)=P_{{\rm st},0}(r,s)[1+\ldots]$.
However, the dot terms turn out to be of order ${\cal O}(1)$ as $N\to\infty$,
which is a sign that this is not the right way to expand.
The reason for this failure is that $\Pst$ is the
exponential of the extensive quantity ${\cal H}_0$; 
one should therefore ask first if this exponential contains 
any corrections of less divergent order in $N$ before attempting to
multiply it by a series of type $[1+\ldots]$.
In the next section we describe how the expansion can be set up
successfully.\\
 
Knowing how to calculate higher order corrections to the
stationary state distribution, although 
certainly of diminishing practical interest,
has a definite theoretical merit.
What we will find in the end 
is that in fact to first order in the expansion detailed balancing continues
to hold, but with respect to a Hamiltonian ${\cal H}^{(1)}(r,s)$
that acquires a first order correction. 
In section \ref{secfirst} we present the solution, to be denoted as
$P^{(1)}_{\rm st}(r,s)$, of the
stationary state to first order. In section \ref{secsecond}
we will show how higher orders can be calculated and
find that from the second order on DB violation appears.
Section \ref{secsecond} 
also provides the demonstration of the correctness of the expansion.


\section{Stationary state $P_{\rm st}(r,s)$ to first order}
\label{secfirst}

We use the upper index `$(1)$' to indicate any quantity correct
up to first order in the expansion.
We will prove that the correct expansion takes the form
\beq
\Pst(r,s) = \Pstupone(r,s)\left[ 1 + q_1(r,s) + q_2(r,s) + \ldots\right],
\label{seriestwo}
\eeq
where 
the $q_k$ $(k=1,2,\ldots)$,
that we will show how to determine later, are of 
of order ${\cal O}(N^{-k/2})$ and
where $\Pstupone(r,s)$, which includes a first order correction 
to the zeroth order result, is explicitly given by
\begin{subequations}\label{deffirstorder}
\beq
\Pstupone(r,s) = {\cal N}^{(1)} \exp
\left( -\frac{{\cal H}^{(1)}(r,s)}{T} \right), 
\label{deffirstP}
\eeq
\beq
\frac{{\cal H}^{(1)}(r,s)}{T} = \frac{{\cal H}_0(r,s)}{T}
-g_0 N \delta\mu(r)\delta\mu(s),
\label{deffirstH}
\eeq
\beq
g_0 = \cosh^2H_0 \tanh\eta K,
\label{deffirstg}
\eeq
\end{subequations}
in which ${\cal N}^{(1)}$ is the appropriate normalization.
The second term on the RHS of (\ref{deffirstH})
is a correction to the zeroth order effective Hamiltonian.
It is ${\cal O}(1)$ for $N\to\infty$ and, since it is proportional to
$g_0$, it vanishes as expected when $\eta=0$.

In order to demonstrate (\ref{seriestwo})-(\ref{deffirstorder})
we split ${\cal W}$ according to
\beq
{\cal W} = \calWupone + \sum_{k=2}^\infty {\cal W}_k\,,
\label{splittwo}
\eeq
where we take for $\calWupone$ the matrix with the factorizing
transition probabilities 
{\it that ensure detailed balancing with respect to\,} ${\cal
  H}^{(1)}$,
and in which the ${\cal W}_k$ will be defined shortly.
Expression (\ref{deffirstorder}) for Hamiltonian ${\cal H}^{(1)}$ shows that a
spin $r_j$ is subject to a total field $H_0 + g_0\delta\mu(s)$.
Hence by analogy to (\ref{defwrj0})
the transition probabilities that enter $\calWupone$ are
\beq
\wupone_j(r;s) = \wGj(r) \times 
\tfrac{1}{2}[ 1 - r_j\tanh\{H_0+g_0\delta\mu(s)\} ].
\label{xwupone}
\eeq
We then have by construction that
\beq
\calWupone\Pstupone =0,
\label{sseqfirstorder}
\eeq
which is the combined zeroth and first order result.
It may be obtained in explicit form
from (\ref{defPst}) by the substitutions 
$w_j \mapsto w_j^{(1)}$ and $\Pst \mapsto \Pstupone$.

A remark on terminology is in place at this point.
Since the zeroth and first order will often be combined, 
we will refer to equation (\ref{sseqfirstorder}) as describing the
`leading order'.
The terms $q_1$, $q_2$, \ldots in the
series (\ref{seriestwo}) will be referred to as `higher order' corrections.


\section{Stationary state to higher orders}
\label{secsecond}

The validity of the expansion procedure
of this section hinges on our being
able to show that the corrections take effectively
the form of the series of $q_k$ in (\ref{seriestwo}), where the terms
are proportional to increasing powers of $N^{-1/2}$.


\subsection{The perturbation series for $P_{\rm st}(r,s)$}
\label{secdefseries}

In order to show that the higher order corrections to $\Pst$ can be
expressed as the series of equation (\ref{seriestwo}),
we must first define the ${\cal W}_k$\, in equation 
(\ref{splittwo}). Let us define $\delta w_j(r;s)$ by
\beq
w_j(r;s) = \wupone_j(r;s) + \delta w_j(r;s).
\label{defdeltaw}
\eeq
Starting from (\ref{defdeltaw}) we  employ the explicit expressions
(\ref{defwrj}) and (\ref{xwupone}) for  $w_j$ and $\wupone_j$, 
respectively, perform a straightforward Taylor expansion in
$\delta\mu$, and still use (\ref{deffield}) to eliminate $m_0$ in
favor of $H_0$. This leads to
\bea
\delta w_j(r;s) &=& \wGj(r) \times \Big[ \tfrac{1}{2}[1-r_j\mu(s)\tanh\eta K] 
- \tfrac{1}{2}[1-r_j\tanh\{H_0+g_0\delta\mu(s)\}] \Big]
\nonumber\\[2mm]
&=& \wGj(r) \times (-\tfrac{1}{2}r_j) \Big[ \{m_0+\delta\mu(s)\}\tanh\eta K
- \tanh\{H_0+g_0\delta\mu(s)\} \Big]
\nonumber\\[2mm]
&=& \wGj(r) \times (-\tfrac{1}{2}r_j) \sum_{k=2}^\infty a_k\,\delta\mu^k(s)
\nonumber\\
&=& \sum_{k=2}^\infty w_{j,k}(r;s),
\label{defwjk}
\eea
where the last equality, supposed to hold term by term in $k$,
defines $w_{j,k}$ and shows that it is of order $N^{-k/2}$.
In the third line of (\ref{defwjk})
the vanishing of the term linear in $\delta\mu$ has of course been
pre-arranged. 
The first two coefficients $a_k$ in that line are given by
\bea
a_2 &=& g_0^2 (1-\tanh^2 H_0)\tanh H_0\,, \nonumber\\[2mm]
a_3 &=& \tfrac{1}{3} g_0^3 (1-\tanh^2 H_0)(1-3\tanh^2 H_0).
\label{xak}
\eea
It becomes clear now that there is a qualitative difference between the
high temperature regime $T\geq \Tc$ where we have
$H_0=0, a_2=0$, and 
\beq
a_3=\frac{1}{3}\tanh^3\eta K,  \qquad T\geq\Tc\,,
\label{xa3}
\eeq
and the low
temperature regime $T<\Tc$ where $H_0>0, a_2>0$. 

We define the matrices ${\cal W}_k$ in expansion
(\ref{splittwo}) in terms of the $w_{j,k}$ by analogy to (\ref{relwW}).
Hence for $T\geq\Tc$ we have that ${\cal W}_2=0$.


\subsection{The higher order equations}
\label{sechigher}

The leading order equation (\ref{sseqfirstorder}) being satisfied,
we now turn to the higher orders. Substitution of (\ref{splittwo}) in 
(\ref{defstst}) and use of (\ref{sseqfirstorder}) leads to an
expansion of which the first term is
\begin{subequations}\label{sseqnextorder}
\beq
\calWupone\Pstupone q_1 + {\cal W}_2\Pstupone = 0, \qquad T<\Tc\,.
\label{ssnextlow}
\eeq
In the high temperature phase the fact that
${\cal W}_2=0$ implies that $q_1$=0 and therefore 
(\ref{ssnextlow}) is replaced by the next term in the expansion,
\beq
\calWupone\Pstupone q_2 + {\cal W}_3\Pstupone = 0, \qquad T\geq\Tc\,.
\label{ssnexthigh}
\eeq
\end{subequations}
Either will be referred to as the `next-to-leading order' equation.
One obtains 
all higher-order equations in explicit form by inserting 
in the full stationary state equation 
(\ref{defPst}) the expansions
(\ref{seriestwo}) for $\Pst(r,s)$ and (\ref{defdeltaw})-(\ref{defwjk}) 
for $w_j(r;s)$.


\subsection{Equation for $T<\Tc$}
\label{seclowT}

By the procedure indicated above we obtain
for the next-to-leading order equation (\ref{ssnextlow}) 
the explicit form
\bea
0 &=& \sum_j \Big[ \wjtwo(r^j;s)\Pstupone(r^j,s) - \wjtwo(r;s)\Pstupone(r,s)
\nonumber\\
&& {}\phantom{XX} + \wjtwo(s^j;r)\Pstupone(r,s^j) - \wjtwo(s;r)\Pstupone(r,s)
\nonumber\\[2mm]
&& {}\phantom{XX} + \wupone_j(r^j;s)\Pstupone(r^j,s)q_1(r^j,s)
   - \wupone_j(r;s)  \Pstupone(r,s)q_1(r,s)
\nonumber\\[2mm]
&& {}\phantom{XX} + \wupone_j(s^j;r)\Pstupone(r,s^j)q_1(r,s^j)
   - \wupone_j(s;r)  \Pstupone(r,s)q_1(r,s) \Big].
\nonumber\\[2mm]
&& {}
\label{ststeqn3}
\eea
We wish to divide (\ref{ststeqn3})
by $\Pstupone(r,s)$ and therefore have to compute 
\beq
\frac{ \Pstupone(r^j,s) }{ \Pstupone(r,s) } \equiv \ee^{-2R_j(r;s)}.
\label{defR}
\eeq
We easily find
\bea
2R_j(r;s) &=& [ \calHupone(r^j,s) - \calHupone(r,s) ]/T
\nonumber\\[2mm]
&=& [ {\cal H}_0(r^j,s) - {\cal H}_0(r,s) ]/T
-g_0N\delta\mu(s)[\delta\mu(r^j)-\delta\mu(r)]
\nonumber\\[2mm]
&=& -2Kr_j(r_{j-1}+r_{j+1}) -2r_j\{H_0+g_0\delta\mu(s)\},
\label{xR}
\eea
where we used (\ref{deffirstorder}) and (\ref{defH0}).
Detailed balancing says that
\beq
\wupone_j(r^j;s)\Pstupone(r^j,s)(r^j,s) = \wupone_j(r;s) \Pstupone(r,s).
\label{detbalW1}
\eeq
Using (\ref{xR}) in the first two lines 
and (\ref{detbalW1})
in the last two lines of (\ref{ststeqn3}) we obtain
\bea
0 &=& \sum_j \Big[ \wjtwo(r^j;s)\ee^{-2R_j(r;s)} - \wjtwo(r;s)
\nonumber\\
&& {}\phantom{XX} + \wjtwo(s^j;r)\ee^{-2R_j(s;r)} - \wjtwo(s;r)
\nonumber\\[2mm]
&& {}\phantom{XX} + \wupone_j(r;s) \{q_1(r^j,s)-q_1(r,s)\}
\nonumber\\[2mm]
&& {}\phantom{XX} + \wupone_j(r;s) \{q_1(r,s^j)-q_1(r,s)\}  \Big].
\label{ststeqn4}
\eea
The expression in the first line of (\ref{ststeqn4}) may be rewritten as
\beq
\begin{split}
\wjtwo(r^j;s)&\ee^{-2R_j(r;s)} - \wjtwo(r;s)\\[2mm]
& = \wGj(r) \times \tfrac{1}{2}[1-r_j\tanh H_0]\times
4r_j\delta\mu^2(s) g_0^2\tanh H_0\,,
\end{split}
\label{x1}
\eeq
of which the first two factors on the RHS
are again exactly $w_{j,0}$\,.
In (\ref{ststeqn4}) 
$\wupone_j$ is of order $N^0$ but contains corrections of higher order in
$N^{-1/2}$. In (\ref{ststeqn4}), 
to leading order in $N^{-1/2}$, we may therefore replace it by its
$N\to\infty$ limit, that is, by $w_{j,0}$ defined by (\ref{defwrj0}).
When we substitute (\ref{x1}) in (\ref{ststeqn4})
and apply to $\wupone_j$ the $N\to\infty$ limit, we obtain the final form of the
equations for the next-to-leading order correction to the stationary state,
\bea
0 &=& \sum_j \Big[
     \wjzero(r) \times 4r_j g_0^2\tanh H_0\,\delta\mu^2(s)
\nonumber\\
&& {}\phantom{XX} + \wjzero(s) \times 4s_j g_0^2\tanh H_0\,\delta\mu^2(r)
\nonumber\\[2mm]
&& {}\phantom{XX} + \wjzero(r)\{ q_1(r^j,s) - q_1(r,s) \}
\nonumber\\[2mm]
&& {}\phantom{XX} + \wjzero(s)\{ q_1(r,s^j) - q_1(r,s) \}
\Big].
\label{ststeqn5a}
\eea


\subsection{Equation for $T \geq \Tc$}
\label{sechighT}

For $T \geq \Tc$ we have $a_2=0$ whence $q_1=0$.
Equation (\ref{ssnexthigh}), when rendered explicit, 
leads to expressions that are identical 
to successively (\ref{ststeqn3}), (\ref{ststeqn4}), and (\ref{ststeqn5a}) 
apart from the substitutions
$q_1\mapsto q_2$ and $\wjtwo\mapsto w_{j,3}$.
In this case $\wjzero(r)=\tfrac{1}{2}\wGj(r)$ where
$\wGj(r)$ is given by (\ref{defwG}), and $\delta\mu=\mu$ since $H_0=m_0=0$.
Hence instead of (\ref{ststeqn5a}) we get
\bea
0 &=& \sum_j \Big[
     \wGj(r) \times 4r_j a_3 \mu^3(s)
\nonumber\\
&& {}\phantom{XX} + \wGj(s) \times 4s_j a_3 \mu^3(r)
\nonumber\\[2mm]
&& {}\phantom{XX} + \wGj(r)\{ q_2(r^j,s) - q_2(r,s) \}
\nonumber\\[2mm]
&& {}\phantom{XX} + \wGj(s)\{ q_2(r,s^j) - q_2(r,s) \}
\Big].
\label{ststeqn5b}
\eea
Finding the solutions of (\ref{ststeqn5a}) and (\ref{ststeqn5b}) 
will be the subject of the next two subsections.
We will first consider the easier case of $T\geq\Tc$ and then the
case $T<\Tc$.


\subsection{Solution for $T \geq \Tc$}
\label{secsolhigh}

We start with the high temperature phase, where equation 
(\ref{ststeqn5b}) applies.
Detailed balancing would be satisfied if the expression under the sum on $j$
were zero, that is, if we had
\bea
q_2(r^j,s) - q_2(r,s) &=& -a_3 r_j \mu^3(s),
\nonumber\\[2mm]
q_2(r,s^j) - q_2(r,s) &=& -a_3 s_j \mu^3(r).
\label{detbal5b}
\eea
It can easily be shown that it is impossible to satisfy these equations.
However, they suggest that we look for a solution $q_2$ of the form
\beq
q_2(r,s) = NC_2a_3[ \mu(r)\mu^3(s) + \mu(s)\mu^3(r) ]
\label{tryb}
\eeq
where only the constant $C_2$ is still adjustable.
The difference $q_2(r^j,s) - q_2(r,s)$ is easy to calculate, but we are
interested only in its leading order.
This leads to
\begin{subequations}\label{difftryb}
\bea
q_2(r^j,s) - q_2(r,s) &=& -2C_2 a_3 r_j [ \mu^3(s) + 3\mu(s)\mu^2(r) ]
+ {\cal O}(N^{-2}),\phantom{XXX}
\label{difftryb1}\\[2mm]
q_2(r,s^j) - q_2(r,s) &=& -2C_2 a_3 s_j [ \mu^3(r) + 3\mu(r)\mu^2(s) ]
+ {\cal O}(N^{-2}).\phantom{XXX}
\label{difftryb2}
\eea
\end{subequations}
It should be noted that whereas (\ref{tryb}) is of order $N^{-1}$,
the differences (\ref{difftryb}) are of order $N^{-3/2}$.
We now need
\beq
\sum_j \wGj(r) \{ q_2(r^j,s) - q_2(r,s) \} =
-2C_2a_3 \left( \sum_j \wGj(r)r_j \right) [ \mu^3(s) + 3\mu(s)\mu^2(r) ].
\label{sumwGj}
\eeq
With the aid of the explicit expression for $\wGj(r)$ one evaluates easily
\beq
\sum_j \wGj(r)r_j = \tfrac{1}{4}(1-\gamma)N\mu(r).
\label{sumb}
\eeq
we see that the equation is
satisfied for $C_2=\tfrac{1}{8}$. Hence from (\ref{tryb}) we get
\beq
q_2(r,s) = \tfrac{1}{24}N (\tanh^3\eta K) [ \mu(r)\mu^3(s) + \mu(s)\mu^3(r) ].
\label{q2final}
\eeq
This is of order $N^{-1}$.
\vspace{5mm}


\subsection{Solution for $T <\Tc$}
\label{secsollow}

In the low-temperature regime equation (\ref{ststeqn5a}) applies.
In order to solve this equation we now postulate
\beq
q_1(r,s) = NC_1b_2 [ \delta\mu^3(r) + \delta\mu^3(s) ]
\label{trya}
\eeq
where $C_1$ is an adjustable constant and
\beq
b_2 = 4a_2/(1-\tanh^2H_0) = 4g_0^2\tanh H_0\,.
\label{defb2}
\eeq
Expression (\ref{trya}) is of order $N^{-1/2}$.
Instead of (\ref{difftryb}) we now have the difference
\beq
q_1(r^j,s) - q_1(r,s) = -6C_1b_2 r_j \delta\mu^2(r) + {\cal O}(N^{-3/2}). 
\label{difftrya}
\eeq
which is of order $N^{-1}$.
The first two lines of (\ref{ststeqn5a}) require that we evaluate
\beq
\sum_j \wjzero(r)r_j = \tfrac{1}{4} \sum_j
[ 1-\tfrac{1}{2}\gamma r_j( r_{j-1}-r_{j+1} )][ 1-r_j\tanh H_0 ]r_j
\label{suma0}
\eeq
Unlike the sum in (\ref{sumwGj}), 
this is not a sum of zero-average random terms.
It will produce a result of order $N$, which we may replace by its average.
This yields
\bea
\sum_j \wjzero(r)r_j &=& \tfrac{1}{4}N[ (1-\gamma)m_0 - 
                                      (1-\gamma a_H) \tanh H_0 ]
\nonumber\\
&\equiv& NG,
\label{suma}
\eea
where the last equality defines $G$ and where $a_H$
is the nearest neighbor spin-spin correlation $\la r_jr_{j+1} \ra$ 
of a 1D Ising chain in a field as described by ${\cal H}_0$ 
[equation (\ref{defH0})].
Expression (\ref{suma}), contrary to its $T\geq \Tc$ counterpart
(\ref{sumb}), has no spin dependence and is therefore
equal for the $r$- and $s$- spins.
The first two lines of (\ref{ststeqn5a}), to be denoted $S_1$, become
\beq
S_1 = 4 NG g_0^2 \tanh H_0\, \big[ \delta\mu^2(s) + \delta\mu^2(r) \big].
\label{xS1}
\eeq
We use (\ref{difftrya}) to write the last two lines of (\ref{ststeqn5a}) as
\bea
S_2 &=& -6C_1b_2 \left[
  \Big( \sum_j\wjzero(r)r_j \Big) \delta\mu^2(r)
+ \Big( \sum_j\wjzero(s)s_j \Big) \delta\mu^2(s)
\right] \nonumber\\[2mm]
&=& -6 NG C_1b_2\,\big[ \delta\mu^2(r) + \delta\mu^2(s) \big].
\label{xS2}
\eea
The stationary state equation (\ref{ststeqn5a}) 
may the be written as $S_1+S_2=0$
and we see that it is satisfied for $C_1=\tfrac{1}{6}$.
\beq
q_1(r,s) = \tfrac{2}{3}N g_0^2 \tanh H_0\, \big[ \delta\mu^3(r) +
\delta\mu^3(s) \big].
\label{q1final}
\eeq


\subsection{Section summary}
\label{sec}

We have studied in the preceding subsections the large-$N$ expansion
of the stationary state distribution $P_{\rm st}(r,s)$
of the infinite velocity CRIC defined in section
\ref{secdefmodel}. 
We have shown, for $T<\Tc$ and $T\geq\Tc$ separately, 
the existence of a series of correction terms $q_k$
that multiplies the leading order result $\Pstupone$ in
(\ref{seriestwo}),
which itself is again composed of a zeroth and a first order contribution.
This expansion also furnishes the necessary proof that
the prefactor $\Pstupone$
represents indeed the `leading order' behavior. 
We have determined explicitly
the first nonzero correction term in this series:
$q_1$ for $T\geq\Tc$ and $q_2$ for $T<\Tc$. 

When looking ahead beyond this leading order correction,
it appears that the $q_k$ (for $k\geq 2$ when $T<\Tc$
and for $k \geq 3$ when $T\geq\Tc$) involve not only $\delta\mu(r)$
and $\delta\mu(s)$, but also energy fluctuations such as
$N^{-1}\sum_j(r_jr_{j+1} - a_H)$, if not longer-range correlations.
Therefore, even though
on the basis of the results of this section
one might be tempted to postulate a general solution of the simple type
$\Pst(r,s) = \Pstupone(r,s)Q(\delta\mu(r),\delta\mu(s))$, 
it is unlikely that the true $P_{\rm st}(r,s)$ is of this form. 


\section{Stationary state averages}
\label{secaverages}

Stationary state averages of observables $A(r,s)$
are averages with respect to $\Pst(r,s)$, so that using (\ref{seriestwo})
and (\ref{deffirstP}) we have
\bea\label{defaverage}
\la A\ra &=& \frac{ \sum_{r,s}A(r,s)\ee^{-{\calHupone}(r,s)/T}
[ 1 + q_1(r,s) + q_2(r,s) + \ldots ] }
{ \sum_{r,s}\ee^{-{\calHupone}(r,s)/T}
[ 1 + q_1(r,s) + q_2(r,s) + \ldots ] }
\nonumber\\[2mm]
&=& \la A \raupone +[\la Aq_\ell\raupone - 
\la A\raupone \la q_\ell\raupone] +\ldots,
\eea
where $\la \ldots \raupone$ indicates an average with weight
$\Pstupone(r,s)$ 
[equation (\ref{deffirstorder})],
the second line results from a straightforward expansion, and
\beq
\ell = \left\{
\begin{array}{ll}
2,\phantom{XXX} & T \geq \Tc\,, \\[2mm]
1,              & T <    \Tc\,,
\end{array}
\right.
\eeq
for the lowest order nonzero terms in the expansion.
Although the $q_k$ are accompanied by increasing powers of
$N^{-1/2}$, the order in $N^{-1/2}$ of each of the terms in the series
(\ref{defaverage}) must be analyzed for each observable $A$
separately. 


\subsection{Integral representation of the partition function} 
\label{secintrepr}

The denominator in the first line of (\ref{defaverage})
is a normalization factor to which we may refer
(although slightly improperly) as the partition function $Z$.
In order to find expressions for the averages $\la \ldots \raupone$ 
in the second line of (\ref{defaverage}), 
we begin by evaluating $Z$ to leading order,
\beq
\Zstupone(K,H_0,g_0) \equiv \sum_{r,s} \ee^{-{\calHupone}(r,s)/T},
\label{defZeff}
\eeq
with $\calHupone$ given by (\ref{deffirstH}) in
which one should substitute (\ref{defH0}) and (\ref{defmurs}).
To this order (\ref{defZeff}) is a true partition function, {\it viz.}
the trace of a Boltzmann factor.
The notation $\Zstupone(K,H_0,g_0)$ is meant to indicate that we wish
to consider this quantity as a function of three independent parameters,
ignoring for the moment expression (\ref{deffirstg}) for $g_0$\,. 
The $r$- and $s$-spins in (\ref{defZeff}) 
may be decoupled by the integral representation
\bea
\Zstupone &=& \frac{N}{\pi g_0}
\int_{-\infty}^{\infty}\!\dd x  \int_{-\infty}^{\infty}\!\dd y\,\,
\ee^{ -g_0^{-1}{N}(x^2+y^2) }            
\nonumber \\[2mm]
&& \times \left[ \ee^{-(x+{\rm i}y)Nm_0 }
\sum_r \ee^{K\sum_j r_j r_{j+1} + (H_0+x+{\rm i}y)\sum_j r_j} \right]
\nonumber\\[2mm]
&& \times \left[ \ee^{-(x-{\rm i}y)Nm_0 }
\sum_s \ee^{K\sum_j s_j s_{j+1} + (H_0+x-{\rm i}y)\sum_j s_j} \right]
\nonumber\\[2mm]
\label{intreprZeff}
\eea
in which $m_0=m(K,H_0)$ follows from (\ref{deffield}) and (\ref{solnH0}).
The two factors in brackets in (\ref{intreprZeff}) 
are seen to be the partition functions $\zeta(K,H_0 + x \pm {\rm i}y)$
of independent standard 
Ising chains in magnetic fields $H_0 + x \pm {\rm i}y$. Hence
\beq
\Zstupone = \frac{N}{\pi g_0} 
\int_{-\infty}^{\infty}\!\dd x  \int_{-\infty}^{\infty}\!\dd y\,
\ee^{ -g_0^{-1}N(x^2+y^2) -2xNm_0 }
\big| \zeta(K,H_0+x+{\rm i}y) \big|^2.
\label{intZeff}
\eeq
We recall that
\beq
\zeta(K,B) \equiv \lambda_+^N + \lambda_-^N\,,
\label{xZeff}
\eeq
where
\beq
\lambda_\pm(K,B) = \ee^K \left[ \cosh B \pm \sqrt{ \sinh^2 B + \ee^{-4K} }
\right].
\label{eigtransf}
\eeq
are the transfer matrix eigenvalues.


\subsection{Stationary point and fluctuations}
\label{secstatpoint}

The $x$ and $y$ integrals in (\ref{intZeff}) are easily evaluated by
the saddle point meyhod,
In the limit of large $N$,
we may neglect in (\ref{xZeff}) 
the exponentially small corrections due to $\lambda_-$
and get from (\ref{intZeff})
\beq
\Zstupone \simeq  \frac{N}{\pi g_0} 
\int_{-\infty}^{\infty}\!\dd x  \int_{-\infty}^{\infty}\!\dd y\,\,
\ee^{ -N {\cal F}(x,y) },
\label{xZeff2}
\eeq
where
\beq
{\cal F}(x,y)= 
g_0^{-1} (x^2+y^2) +2xm_0 
- \log\big|\lambda_+(K,H_0+x+{\rm i}y)\big|^2 
\label{defcalF}
\eeq
Let $(x^*,y^*)$ denote the stationary point of the integration in 
(\ref{xZeff2}).
The stationary point equations ${\cal F}_x={\cal F}_{y} =0$ 
can be expressed as
\beq
g_0^{-1}(x^*\pm{\rm i}y^*) = m(K,H_0+x^*\mp{\rm i}y^*) - m_0\,,
\label{statpoint}
\eeq
with the magnetization
$m(K,B)=\lambda_+^{-1}(K,B){\partial\log\lambda_+(K,B)}/{\partial B}$
given by (\ref{xmKz}).
For reasons of symmetry the stationary point 
must have $y^*=0$.
This reduces (\ref{statpoint}) to the single real equation
\beq
g_0^{-1}x^* = m(K,H_0+x^*) - m(K,H_0),
\label{statpointreal}
\eeq 
where we used that $m_0=m(K,H_0)$ [equation (\ref{relconst})].
Equation (\ref{statpointreal}) 
has for all $H_0$ the obvious solution $x^*=0$.
We investigate the stability of the stationary point $(x^*,y^*)$
by calculating the matrix of second derivatives, 
\bea
&&{\cal F}_{xx}^* = 2[g_0^{-1} - \chi(K,H_0)], \qquad
  {\cal F}_{yy}^* = 2[g_0^{-1} + \chi(K,H_0)], \nonumber\\[2mm]
&&{\cal F}_{xy}^* = {\cal F}^*_{yx} = 0, 
\label{calFsecond}
\eea
where the asterisk indicates evaluation in the stationary point
and where $\chi(K,B)=\partial m(K,B)/\partial B$ is the magnetic 
susceptibility. We obtain 
the eigenvalues ${\cal F}_{xx}^*$ and ${\cal F}_{yy}^*$ explicitly
by substituting in (\ref{calFsecond}) for $g_0$
the expressions (\ref{deffirstg}) and for 
$\chi$ the expression
\beq
\chi(K,B) = 
\frac{\ee^{-4K}\cosh B}{ \left( \sinh^2 B + \ee^{-4K} \right)^{3/2} }\,,
\label{defchiKB}
\eeq
where (\ref{xmKz}) has been used. This yields
\beq
{\cal F}_{xx,yy}^* =
\left\{
\begin{array}{ll} 
\dfrac{ 2(\ee^{-2K} \mp \tanh\eta K) }{\ee^{-2K}\tanh\eta K}, & T>\Tc\,, \\[4mm]
\dfrac{ 2 (1-\tanh^2\eta K) (\tanh^2\eta K \mp {\ee}^{-4K}) }
{ (1-{\ee}^{-4K}) \tanh^3\eta K }\,, & T<\Tc\,,
\end{array}
\right.
\label{xchilow}
\eeq
in which the upper (lower) sign refers to the $xx$ (to the $yy$) derivative.
It can be seen that ${\cal F}_{yy}^*$ is positive for all
temperatures, but that ${\cal F}_{xx}^*$\,,
which is positive in both the high and
the low-temperature phase, vanishes as $T\to\Tc$.
Hence for all $T\neq\Tc$ the stability is ensured by the quadratic terms
in the expansion of ${\cal F}(x,y)$ around the stationary point.


\subsection{Free energy}

We are now in a position to calculate various physical quantities of interest.
The first one will be the 
{\it interaction\,} free energy per spin between the two
chains which (divided by $T$)
will be called $F_{\rm int}$. It will turn out to have an expansion
\beq
F_{\rm int} = F_{\rm int}^{(0)} + N^{-1}f_{\rm int} +\ldots
\label{formFint}
\eeq
To show this we pursue the calculation of $\Zstupone$
begun in (\ref{xZeff2}). We there substitute the expansion
\beq
{\cal F}(x,y) = {\cal F}^* + \tfrac{1}{2}{\cal F}_{xx}^* x^2
                              + \tfrac{1}{2}{\cal F}_{yy}^* y^2 + \ldots\,.
\label{expcalF}
\eeq
We can then carry out the integrations in (\ref{xZeff2}) by the saddle point
method and find that only the quadratic terms in
(\ref{expcalF}) contribute. 
The result has the form
\beq
\Zstupone \simeq \ee^{-N{\cal F}^* - f_{\rm int}}
\left[1+{\cal O}(N^{-1})\right]
\label{resZeff}
\eeq  
where
\beq 
N{\cal F}^*=N{\cal F}(0,0)=-2N\log\lambda_+(K,H_0)
\label{xcalFstar}
\eeq
and
\beq
f_{\rm int}(K,\eta) = \tfrac{1}{2}\log\left[ 1-g_0^2\chi^2(K,H_0) \right].
\label{xfint}
\eeq
Here ${\cal F}^*$ is the free energy (divided by $T$) 
of two independent Ising chains in an effective
field $H_0$.
Since $H_0$ is proportional to the coupling $\eta K$
between the chains,  
the field dependent
part of ${\cal F}^*$ actually represents the bulk interaction free energy 
$NF^{(0)}_{\rm int}$ between the chains, that is,
\beq
NF^{(0)}_{\rm int}(K,\eta) = 
\left\{
\begin{array}{ll}
0, & T\geq\Tc\,,\\[2mm]
-2N\log\left( \dfrac{\lambda_+(K,H_0)}{\lambda_+(K,0)} \right), & T<\Tc\,; 
\end{array}
\right.
\label{xxFint}
\eeq
and furthermore $f_{\rm int}(K,\eta)$
is a residual interaction free energy between them which remains of
order $N^0$ as $N\to\infty$.
The energy that one drives from it has a cusp singularity and hence
the exponent $\alpha=0$ \cite{Hucht09}.

Beyond this leading order result
we obtain $f_{\rm int}$ explicitly in terms of the two system parameters
$K$ and $\eta$ by substituting in (\ref{xfint}) the expressions for 
$g_0$ and $\chi$ given in (\ref{deffirstg}) and (\ref{defchiKB}), 
respectively, and (when $T<\Tc$) eliminating $H_0$. The result is that
\bea
f_{\rm int}(K,\eta) &=& \left\{
\begin{array}{ll}
\tfrac{1}{2}\log\left( 1-{\ee}^{4K}\tanh^2\eta K \right), & T >\Tc\,,\\[2mm]
\tfrac{1}{2}\log\left( 1-{\ee}^{-8K}\tanh^{-4}\eta K \right), & T <\Tc\,.
\end{array}
\right.
\label{xxfint}
\eea
In view of (\ref{xxFint}) 
we see that $F_{\rm int}$ has a linear cusp at $T=\Tc$, and
(\ref{xxfint}) shows that
$f_{\rm int}$ diverges logarithmically for
$T\to\Tc$. In spite of this weak divergence,
the finite size correction $f_{\rm int}$
to the interaction free energy $F_{\rm int}$
also conforms the classical specific heat exponent 
$\alpha=0$.


\subsection{Finite size scaling of the free energy near $\Tc$}
\label{fsscaling}

We will show how our approach allows for
finding the finite size scaling functions.
By the way of an example we consider the singular part of the
free energy. For $T\to\Tc$ the quantity $f_{\rm int}$ diverges
due to the second order derivative ${\cal F}_{xx}^*$ becoming zero.
In order
for the integral (\ref{xZeff2}) combined with (\ref{expcalF})
to converge at $T=\Tc$, we have to include higher order
terms in the expansion (\ref{expcalF}).
We will write
\beq
{\cal F}(x,y) = {\cal F}^* 
                + \tfrac{1}{2}{\cal F}_{xx}^* x^2
                + \tfrac{1}{2}{\cal F}_{yy}^* y^2 
                + \tfrac{1}{6}{\cal F}_{xxx}^* x^3
                + \tfrac{1}{24}{\cal F}_{xxxx}^* x^4 + \ldots
\label{expcalFTc}
\eeq
and will argue below
that near $\Tc$ the terms not exhibited explicitly in this
series are of higher order%
\footnote{Terms with an odd number of $y$ derivations vanish by symmetry.}.
In order to find the coefficients in (\ref{expcalFTc}
we perform a
straightforward derivation of (\ref{defcalF}) and set $x^*=y^*=0$.
We then define
\beq
\epsilon=\frac{T-\Tc}{\Tc}=-\frac{K-\Kc}{\Kc}
\label{defeps}
\eeq
which, in the vicinity of $\Tc$\,, leads to
\beq
H_0 = \Bc\,\epsilon^{1/2} + {\cal O}(\epsilon)
\label{xKHepsilon}
\eeq
where from (\ref{solnH0}) we have 
\beq
\Bc^2 = \left\{
\begin{array}{ll}
0, & T>\Tc\,,\\[2mm]
2\ee^{-2\Kc} 
\big( \eta + 1/\sinh 2\Kc \big)\Kc\,,
& T<\Tc\,.
\end{array}
\right.
\label{defBc}
\eeq
When using (\ref{xKHepsilon}) in the coefficients found above
we obtain
\bea
{\cal F}^*_{xx} &=& a_{\pm}\epsilon + {\cal O}(\epsilon^2), 
\nonumber\\[2mm]
{\cal F}^*_{yy} &=& 4\ee^{2\Kc} + {\cal O}(\epsilon),
\nonumber\\[2mm]
{\cal F}^*_{xxx} &=& b_\pm(-\epsilon)^{1/2} + {\cal O}(\epsilon^{3/2}),
\nonumber\\[2mm]
{\cal F}^*_{xxxx} &=& c +{\cal O}(\epsilon),
\label{coefficients}
\eea
where
\bea
a_\pm &=& \left\{
\begin{array}{r}
2\\[2mm]
4
\end{array}
\right\} \big( \ee^{4\Kc}-1 \big) 
\big( \eta + 1/\sinh 2\Kc \big)\Kc\,,
\qquad
\begin{array}{l}
T>\Tc\,,\\[2mm]
T<\Tc\,.
\end{array}
\nonumber\\[2mm]
b^2_\pm &=& \left\{
\begin{array}{ll}
0, & T>\Tc\,,
\nonumber\\[2mm]
4 \big( 3\ee^{4\Kc}-1 \big) 
\big( \eta + 1/\sinh 2\Kc \big)\Kc\,,\quad{}
& T<\Tc\,.
\end{array}
\right.\\[2mm]
c &=& 6\ee^{2\Kc},
\label{defabc}
\eea
We substitute the explicit expressions (\ref{coefficients}) in
(\ref{expcalFTc}) and use that expansion in the integral
(\ref{xZeff2}). When we introduce the scaled variables of integration
$u$ and $v$ defined by
\beq
x=N^{-1/4}u, \qquad y=N^{-1/2}v,
\label{scaledvar}
\eeq
as well as the scaling variable
\beq
\tau=\epsilon N^{1/2},
\label{deftau}
\eeq
the factor $N$ disappears from the exponential. After carrying out
the Gaussian integration on $v$ we get  
\beq
\Zstupone \simeq  \ee^{-N{\cal F}^*} \times
\frac{N^{1/4}\ee^{\Kc}}{\sqrt{2\pi}} 
\,{\cal Z}(\tau),
\label{xZKc}
\eeq
valid in the scaling limit $N\to\infty$, $T\to\Tc$ with $\tau$ fixed, 
and where ${\cal Z}$ is the scaling function
\beq
{\cal Z}(\tau) = \int_{-\infty}^{\infty}\!\dd u\,  
\exp\left[-\tfrac{1}{2} a_\pm|\tau| u^2
  -\tfrac{1}{6} b_\pm(-\tau)^{1/2}u^3 
  -\tfrac{1}{24} cu^4 \right].
\label{defcalZ}
\eeq
It is of a type that occurs 
standardly in problems with mean field type critical
behavior; they have been studied recently by  Gr\"uneberg and Hucht 
\cite{GrunebergHucht04}.
It has the limiting behavior
\beq
{\cal Z}(\tau) \simeq
\left\{
\begin{array}{ll}
{\cal Z}(0) \equiv
\int_{-\infty}^\infty\!\dd u\,\ee^{-cu^4/24}, & \tau\to 0, \\[4mm]
\left( \dfrac{2\pi}{a_\pm}|\tau| \right)^{1/2}\,,& \tau\to\pm\infty.
\end{array}
\right.
\label{xcalZlimit}
\eeq
Upon combining (\ref{resZeff}) and (\ref{xZKc}) we find that
\bea
f_{\rm int}(K,\eta) = -\tfrac{1}{4}\log N 
-\log {\cal Z}(tN^{1/2}) 
-\tfrac{1}{2}\log\left( \dfrac{\ee^{2\Kc}}{2\pi} \right) + \ldots,
\label{scalingfint}
\eea
again valid in the scaling limit,
and where the dots stand for terms that vanish as $N\to\infty$.
It follows, in particular, that equation (\ref{xxfint}) may 
now be completed by
\beq
f_{\rm int}(\Kc,\eta) \simeq \tfrac{1}{4}\log N + \log{\cal Z}(0) + \ldots, 
\qquad T=\Tc\,, \quad N\to\infty,
\label{xxfintTc}
\eeq
where the dots stand for terms that vanish as $N\to\infty$.


\subsection{Susceptibilities}
\label{secsusceptibilities}

Of primary interest are the correlations between the fluctuations of
the magnetizations in the two chains. We set as before $\delta\mu=\mu-m_0$.
The general expression that we
will study here is
\bea
\chi_{k\ell}&\equiv& \la \delta\mu^k(r)\delta\mu^\ell(s)\ra
\nonumber\\[2mm]
&=&\la \delta\mu^k(r)\delta\mu^\ell(s)\raupone + \ldots\,
\label{defchimn}
\eea
where the dots in the last line, obtained according to (\ref{defaverage}), 
represent higher order terms.
Special cases that we will consider
are the cross-chain susceptibility $\chi_{\rm int}$  
and the single-chain susceptibility $\chi_{\rm sin}$, defined as
\begin{subequations}\label{chispecial}
\bea
\chi_{\rm int} &=& N\chi_{11}=N\la\delta\mu(r)\delta\mu)(s)\ra, 
\label{chispecialint}\\[2mm]
\chi_{\rm sin} &=& N\chi_{20}=N\la\delta\mu^2(r)\ra,
\label{chispecialsin}
\eea 
\end{subequations}
in which, of course, the latter is also equal to $\chi_{02}$ by symmetry.


\subsubsection{Cross-susceptibility}
\label{seccrosssusc}

We first consider the correlations between the
fluctuating magnetizations of the two chains.
The cross-susceptibility $\chi_{\rm int}$
is the quantity most characteristic of these correlations.
From equations (\ref{deffirstH}) and (\ref{defZeff}) it is clear 
that $\chi_{\rm int}=\partial\log\Zstupone/\partial g_0$
where the derivative has to be evaluated at fixed $K$ and $H_0$,
considering $g_0$ as an independent parameter in (\ref{intreprZeff}).
Doing the calculation for $\Zstupone$ given by (\ref{resZeff}),
(\ref{xcalFstar}), and (\ref{xfint}), we observe that 
${\cal F}^*={\cal F}(0,0)$ is independent of $g_0$ so that
\bea
\chi_{\rm int}(K,\eta) =\frac{\partial f_{\rm int}}{\partial g_0} 
&=& \frac{g_0\chi^2}{1-g_0^2\chi^2} \nonumber\\[2mm]
&=& \left\{
\begin{array}{ll}
\dfrac{ \tanh\eta K }{ \ee^{-4K}-\tanh^2\eta K } & T>\Tc\,,\\[4mm]
\dfrac{ {\rm e}^{-8K} (1-\tanh^2\eta K) }
      {  (\tanh\eta K) (1-{\rm e}^{-4K}) 
        (\tanh^4\eta K-{\rm e}^{-8K})} & T<\Tc\,.
\end{array}
\right.
\label{reschiint}
\eea
For $T\to\Tc$ this quantity diverges as 
$|T-\Tc|^{-\gamma_{\rm int}}$ 
with 
$\gamma_{\rm int}=1$. 
It is a signal that at $T=\Tc$  this correlation scales with another
power of $N$. A scaling function for $\chi_{\rm int}$ may be derived from
the one for $f_{\rm int}$\,, but
we will not try to be exhaustive.

Since at speed $v=\infty$ all index pairs $(i,j)$ are equivalent,
the correlations between the $r$- and the $s$-spins are given by 
\beq
\langle r_i s_j \rangle -m_0^2 =N^{-1}\chi_{\rm int}(K,\eta). 
\label{xrisjav}
\eeq


\subsubsection{Single-chain chain susceptibility}
\label{secsinglesusc}

The single-chain susceptibilities $\chi_{\rm sin}$ 
is defined in equation (\ref{chispecial}).
Let us now consider the general expression (\ref{defchimn}) 
for $\chi_{k\ell}$\,, for which the appropriate
approach differs slightly from that of the preceding subsection.
One may generate insertions $\delta\mu^k(r$ [or $\delta\mu^\ell(s)$]  
in the integral (\ref{intZeff}) by
passing from $x$ and $y$ to the two
independent variables $z=(x+{\rm i}y)$ and $\bz=(x-{\rm i}y)$ and 
letting $N^{-k}\partial^k/\partial z^k$ 
[or $N^{-\ell}\partial^\ell/\partial \bz^\ell$]
act on $\ee^{-2zNm_0}Z(K,H_0+z)$ [or on $\ee^{-2\bz Nm_0}Z(K,H_0+\bz)$].
We find, using (\ref{xZeff}) and neglecting again the effect of $\lambda_-$
which is exponentially small in $N$,
\beq
N^{-k}\frac{\partial^k}{\partial z^k}\,\,
\big[ \ee^{-zNm_0}Z(K,H_0+z) \big] = J_k(z)\,Z(K,H_0+z),
\label{partialkz}
\eeq
in which
\bea
J_0(z)&=& 1, \nonumber\\[2mm]
J_1(z)&=& \tm-m_0\,, \nonumber\\[2mm]
J_2(z) &=& (\tm-m_0)^2+N^{-1}\tchi, \nonumber\\[2mm]
J_3(z) &=& (\tm-m_0)^3 +3N^{-1}(\tm-m_0)\tchi +N^{-2}\tchi', \nonumber\\[2mm]
J_4(z) &=& (\tm-m_0)^4 + 6N^{-1}(\tm-m_0)^2\tchi + 4N^{-2}(\tm-m_0)\tchi'
\nonumber\\[2mm]
       & & +3N^{-2}\tchi^2 + N^{-3}\tchi^{\prime\prime},
\label{defJkz}
\eea
where, in this formula, 
we abbreviated $\tm=m(K,H_0+z)$ and $\tchi=\chi(K,H_0+z)$ 
[see equations (\ref{xmKz}) and (\ref{defchiKB})] 
in order to emphasize the $z$ dependence of these quantities,
and where the
primes on $\tchi$ stand for differentiations with respect to $H_0$.
Equations (\ref{partialkz}) and (\ref{defJkz}) of course have 
counterparts obtained by 
letting $r\mapsto s$, $k\mapsto\ell$ and $z\mapsto\bar{z}$. 
When (\ref{partialkz}) is substituted in (\ref{defchimn}) we obtain
\beq
\chi_{k\ell} = \la J_k(z)J_{\ell}(\bar{z})\raupone + \ldots,
\label{xchimnJ}
\eeq
where the dots stand for higher-than-leading order terms in the
$N^{-1}$ expansion.

By virtue of equations (\ref{xchimnJ}) and (\ref{defJkz}) 
it follows that
\bea
\chi_{20} &=& \la J_2(z)\raG \nonumber\\[2mm]
                  &=& \la \big( m(K,H_0+z)-m_0 \big)^2 \ra  
              + N^{-1}\la\chi(K,H_0+z)\ra 
\label{x2chi20}
\eea
We now expand $m$ and $\chi$ for small $z$ 
anticipating that upon integration 
with weight $\exp(-N{\cal F})$ each factor $z^2$ 
will, to leading order, 
produce a factor $N^{-1}$. After multiplication by $N$ this yields
\beq
\chi_{\rm sin}(K,\eta) 
= N\chi^2(K,H_0)\la z^2\raG + \chi(K,H_0) +{\cal O}(N^{-1}).
\label{x2chi20bis}
\eeq
Anticipating again that each factor $z$ or $\bz$ will produce a
factor $N^{-1/2}$, we see that all terms exhibited explicitly on the
right hand sides in (\ref{xavJ}) are of order $N^{-1}$.
We have replaced the averages $\la\ldots\raupone$, which are with
respect to $\exp(-N{\cal F}(x,y)$,
by averages $\la\ldots\raG$ in which ${\cal F}(x,y)$ of equation (\ref{defcalF})
is replaced with the Gaussian terms in its expansion, shown in (\ref{expcalF}).

Upon using in (\ref{x2chi20}) the explicit evaluations
\bea
\la z^2 \raG &=& \la x^2 \raG - \la y^2 \raG \nonumber\\[2mm]
&=& \frac{1}{N}\left( \frac{1}{{\cal F}_{xx}^*} - \frac{1}{{\cal F}_{yy}^*}
\right) =
\frac{g_0^2 \chi}{N(1-g_0^2\chi^2)}\,, \qquad T\neq\Tc\,.
\label{xavzz}
\eea
we arrive at
\beq
\chi_{\rm sin}(K,\eta) 
= \frac{\chi}{1-g_0^2\chi^2}\,, \qquad T\neq \Tc\,,
\label{x3chi20}
\eeq
valid in the limit $N\to\infty$.
Hence the in-chain susceptibility $\chi_{\rm sin}$
is equal to the susceptibility of the 1D Ising model enhanced
by a factor $(1-g_0^2\chi^2)^{-1}$ due to
the presence of the other chain.

Using expressions (\ref{deffirstg}) and (\ref{defchiKB}) 
for $g_0$ and $\chi$, respectively, 
we may render (\ref{x3chi20})
explicit in terms of $K$ and $\eta$ and get 
\beq
\chi_{\rm sin}(K,\eta) = 
\left\{
\begin{array}{ll}
\dfrac{ \ee^{-2K} }{ \ee^{-4K}-\tanh^2\eta K }\,, 
\phantom{XXX} & T>\Tc\,,
\\[4mm]
\dfrac{ {\rm e}^{-4K}  (\tanh\eta K) (1-\tanh^2\eta K) }
      { (1-{\rm e}^{-4K}) (\tanh^4\eta K-{\rm e}^{-8K})}\,, &  T<\Tc\,.
\end{array}
\right.
\label{x3chi20bis}
\eeq
For $T\to\Tc$ the susceptibility $\chi_{\rm sin}$
diverge as $(T-\Tc)^{-\gamma}$ with, again,  the classical critical
exponent $\gamma=1$. 
For $\eta=0$ (whence $\Tc=0$) the first one of
equations (\ref{x3chi20bis}) reduces to
the standard susceptibility of the zero field 1D Ising chain.

In agreement with the symmetry of the problem, $\chi_{\rm int}$ is odd
and $\chi_{\rm sin}$ is even in $\eta$. 
Both above and below $\Tc$ one easily verifies that
in agreement with Schwarz's inequality
we have $\chi_{\rm int}/\chi_{\rm sin}\leq 1$. 


\subsection{Spontaneous magnetization}
\label{secspontmagn}

For $T\geq\Tc$ symmetry dictates that the magnetization
$\la\mu(r)\ra$ and $\la\mu(s)\ra$ are zero to all orders.
However, for $T<\Tc$ the magnetization
$\mu(r)=N^{-1}\sum_{j=1}^Nr_j$ has, to leading order, a Gaussian
probability distribution of width $N^{-1/2}$ around $m_0(K,H_0)$.
As a consequence
$\la\delta\mu(r)\ra$ vanishes to order $N^{-1/2}$.
However, to order $N^{-1}$ there appear
nonzero corrections terms to $\la\mu(r)\ra$. 
As an application of equation (\ref{defaverage}) we calculate in this
subsection these correction terms.

Upon using (\ref{defaverage}) for the spacial case $A=\delta\mu(r)$
and inserting in it the explicit expression
(\ref{q1final}) for $q_1$ we obtain
\beq
\la\delta\mu(r)\ra = \la\delta\mu(r)\raupone + \tfrac{2}{3}Ng_0^2\tanh H_0
       \left[ \la J_4(z)\raupone \,+\, 
              \la J_1(z)J_3(\bz)\raupone \right].
\label{xmurav}
\eeq
When substituting (\ref{defJkz}) in the second term of
(\ref{xmurav}) we see that we need
\bea
\la J_4(z)\raupone &=& \chi^4\la z^4\raG +6N^{-1}\chi^3\la z^2\raG
                       +3N^{-2}\chi^2 + {\cal O}(N^{-5/2}),
\nonumber\\[2mm]
\la J_1(z)J_3(\bz)\raupone &=& \chi^4\la z\bz^3\raG 
 +3N^{-1}\chi^3\la z\bz\raG + {\cal O}(N^{-5/2}).
\label{xavJ}
\eea
We have replaced the averages $\la\ldots\raupone$
by averages $\la\ldots\raG$ for the same reasons as in the preceding
subsection. Taking into account again that each factor $z$or $\bar{z}$
brings in a power $N^{-1/2}$, we see that all terms explicitly
exhibited on the right hand sides of equations (\ref{xavJ}) 
are of the same order in $N$, namely
${\cal O}(N^{-2})$.  
The Gaussian averages are easily calculated and we are led to
\beq
\la J_4(z)\raupone  \,+\, \la J_1(z)J_3(\bz)\raupone =
\frac{ 3\chi^3(\chi+g_0) }{ N^2(1-g_0^2\chi^2)^2 } + {\cal O}(N^{-5/2}).
\label{finJav}
\eeq
We should now evaluate the first term on the right hand side of
(\ref{xmurav}), namely 
\beq
\la\delta\mu(r)\raupone = \la J_1(z)\raupone = \chi\la z\raupone.
\label{zavfull1}
\eeq
The Gaussian average $\la z\raG$ 
vanishes on account of symmetry. However, when
the third order terms in the Taylor expansion (\ref{expcalF}) 
of ${\cal F}(x,y)$
are kept and we expand these 
we get after a straightforward
calculation that we will not reproduce here,
\bea
\la\delta\mu(r)\raupone &=& \
\tfrac{1}{3}N\chi\chi'\big[ \la x^4\raG - 3\la x^2y^2\raG \big]
\nonumber\\[2mm]
&=& N\chi\chi'\la x^2\raG\big[ \la x^2\raG - 
\la y^2\raG \big] +{\cal O}(N^{-2})
\nonumber\\[2mm]
&=&\frac{g_0^3\chi^2\chi'}{2N(1-g_0\chi)^2(1+g_0\chi)} + {\cal O}(N^{-2}).
\label{zavfull2}
\eea
The final result for $\la\delta\mu(r)\ra$ is obtained by substitution
of (\ref{zavfull2}) and (\ref{finJav}) in (\ref{xmurav}). 
We see that $\la\delta\mu(r)\ra$ has two contributions of order $
N^{-1}$. The contribution $\la\delta\mu(r)\raupone$ comes from the
effective leading order Hamiltonian $\calHupone$.
The second contribution accompanies the violation of detailed
balancing symmetry and is therefore essentially a non-thermodynamic effect.


\subsection{Pair correlation function}
\label{secpaircorrelation}

It is of interest to study the pair correlation 
\beq
g_N(\ell)\equiv\la r_jr_{j+\ell}\ra
\label{defgNell}
\eeq
in a single chain. To that end we 
consider again expansion (\ref{defaverage}), 
now with $A=r_jr_{j+\ell}$. Its first term may be written
\beq
\gupone_N(\ell) = {\Zstell}/{\Zstupone}
\label{defgNell1}
\eeq
where $\Zstell$ is given by (\ref{intreprZeff}) but with an
insertion $r_jr_{j+\ell}$ in the sum on $r$. Equivalently, $\Zstupone$
is given by the
same integral as (\ref{xZeff2}) but with an insertion
$\tgupone_N(\ell;K,H_0+z)$, this quantity being
the pair correlation of the 1D Ising chain 
in a field $H_0+x+{\rm i}y$.
Evaluation by means of the standard transfer matrix method yields 
\beq
\tgupone_N(\ell;K,H_0+z) = m^2(K,H_0+z)
+\,\frac{\ee^{-4K}\,\tLambda^\ell(K,H_0+z)}{\sinh^2(H_0+z)+\ee^{-4K}}\,,
\label{xtgamma}
\eeq
well-known in the case $z=0$, in which we defined
$\tLambda = {\lambda_-}/{\lambda_+}$\,,
where the tilde serves as a reminder of the $z$ dependence,
and where contributions exponentially small in $N$ have again been neglected.
In order to obtain the desired physical correlation function 
$g_N(\ell)$ of this system we now have to average (\ref{xtgamma})
with an appropriately normalized weight 
$\exp\big[ -N{\cal F}(x,y) \big]$.

We will consider this quantity in the high-temperature regime $T>\Tc$
where $H_0=0$.
Knowing that $z$ is of order $N^{-1/2}$ we expand 
(\ref{xtgamma}) for small $z$, which gives
\bea
\tgupone_N(\ell;K,H_0+z) &=& \nonumber\\[2mm]
\ee^{4K}z^2 &+& (\tanh^\ell K)(1-\ee^{4K}z^2)
\exp\left( -(\ee^{-4K}+\ee^{2K}\ell)z^2 \right) + {\cal O}(N^{-2}).\\
&&
\label{xtgammaexp}
\eea
To leading order the average on $z$ may be carried out with the weight
$\exp\big[ -N{\cal F}(x,y) \big]$ in which the expansion ${\cal F}$
is limited to its quadratic terms.
Straightforward calculation yields
\beq
g_N(\ell)= \tanh^\ell K 
+ (1-\tanh^\ell K)\frac{g_0^2\chi^3}{1-g_0^2\chi^2}\,N^{-1} + {\cal O}(N^{-2}),
\qquad T>\Tc\,,
\label{resgNell2}
\eeq
valid for $N\to\infty$ at fixed $\ell$,
where as before $\chi$ stands for the susceptibility $\chi(K,0)=\ee^{2K}$
of the 1D Ising chain and where $g_0=\tanh\eta K$.
In the scaling limit $\ell,N\to\infty$ with a fixed ratio one obtains
\beq
g_N(\ell) \simeq 
(\tanh^\ell K)  \phi(\ell N^{-1}) + \frac{g_0^2\chi^3}{1-g_0^2\chi^2}\,N^{-1},
\quad T>\Tc\,,\quad \ell,N\to\infty, 
\label{resgNell1}
\eeq
in which each of the two terms is valid
up to corrections of relative order $ N^{-1}$
and in which $\phi$ is the scaling function defined by
\beq
\phi^2(x)=\frac{1-g_0^2\chi^2}
{1-g_0^2\chi^2 +2g_0^2\chi^2 x}\,.
\label{defphi}
\eeq
We observe the noncommutativity
\beq
\lim_{N\to\infty} \sum_{\ell=-N/2+1}^{N/2}g_N(\ell)
\neq \sum_{\ell=-\infty}^{\infty}\lim_{N\to\infty}g_N(\ell).
\label{noncomm}
\eeq
The right hand side of this inequality is equal to $\chi(K,0)$
whereas the right hand side is equal to 
$\chi(K,0)+\chi_{\rm int}(K,\eta)$.

We conclude by noting that the pair correlation function may also be studied
to higher order in $N^{-1}$ in the low-temperature regime. For $T<\Tc$
the fluctuations of the magnetic field $z$ are asymmetric and greater
care is required. We will not include such a calculation here.


\section{Traffic model}
\label{sectraffic}

Motivated by an interest very different from that of references 
\cite{Hucht09,Kadauetal08}
we recently introduced a new traffic model describing
vehicles that may overtake each other
on a road with two opposite lanes \cite{ARHS10}.
That work shows the appearance of a phase transition when the traffic
intensity, supposed equal on the two lanes, attains a critical value.
Above the critical intensity the symmetry between the two traffic lanes is
broken: one lane has dense and slow, the other one dilute
and fast traffic.
The study of reference \cite{ARHS10}
invoked a mean-field-type assumption that 
couples the velocity of a vehicle in a given lane to the
{\it average\,} of the vehicle velocities in the opposite lane.
This assumption was justified by the argument that 
a vehicle in one lane 
encounters, in the course of time, all vehicles in the opposite lane. 
Although there is no one-to-one correspondence between the two models,
they share essentially the same features, as may be seen as
follows.
For $J_2<0$ the two chains of the CRIC studied here
have opposite spontaneous
magnetizations; up-spins may then be regarded as the vehicles of the
traffic problem; they will be denser in one chain (traffic lane) than in
the other. 
The CRIC is more amenable to analysis than the traffic model.
It was shown analytically \cite{Hucht09,Kadauetal08} 
that the CRIC phase transition
disappears when $v$ is finite.
Our simulations \cite{ARHS11} of the traffic model have shown,
nevertheless, that this problem is
close to the critical point $v=\infty$. This explains the
critical-point-like phenomena
that we observed, namely
fluctuations that last longer than the simulation time. 


\section{Conclusion}
\label{secconclusion}

We have considered in this paper the nonequilibrium steady state (NESS)
of a model consisting of two counter-rotating interacting Ising
chains introduced by Kadau {\it et al.} \cite{Kadauetal08} and by
Hucht \cite{Hucht09}. The model is
related to a road traffic model studied earlier by ourselves \cite{ARHS10}.
Its dynamics is governed by a master equation parametrized by two
interaction constants $J/T$ and $\eta$.
The model has a phase transition, known to be of mean field type,
at a critical temperature $T=\Tc$\,.

Starting from the master equation
we have shown that in the limiting case
of a relative velocity $v=\infty$ of the two chains,
the stationary state distribution $P_{\rm st}$ may be studied 
in an expansion in powers of the inverse system size $N^{-1}$.
Knowing this distribution we have calculated, also as 
expansions in $N^{-1}$, of averages of physical interest:
the interaction free energy between the chains, the in-chain and
cross-chain susceptibilities, the correlation function (for $T>\Tc$),
and the spontaneous magnetization (for $T<\Tc$). 
We have shown how
near criticality scaling functions may be explicitly calculated. 

Whereas to leading order the force exerted by
one chain on the other is that of 
an effective magnetic field $H_0$,
the $N^{-1}$ expansion requires that we take into account the
fluctuations of this field around its average.
It then appears that to leading order the dynamics obeys detailed
balancing with respect to an effective Hamiltonian, as was found by
Hucht \cite{Hucht09}, but that to higher order in the expansion the
detailed balancing is violated.

In this work we have addressed many different, albeit interrelated,
aspects of the finite-size CRIC. 
We have not tried to be exhaustive and have not considered, for example, 
energy dissipation.
Similarly, the parallel problem with open boundary conditions
has been left aside.
We hope that the results of this work will be helpful in guiding the study,
which we believe to be worthwhile,
of the finite-velocity ($v<\infty$) version of the model.


\section*{Acknowledgments}

The author thanks C\'ecile Appert-Rolland and Gr\'egory Schehr 
(Orsay, France)
and Samyr J\'acobe (UFRN, Natal, Brazil) for
discussions on and around the subject of this work.
 

\appendix

\end{document}